# FEATURE EXTRACTION AND FEATURE SELECTION: REDUCING DATA COMPLEXITY WITH APACHE SPARK

Dimitrios Sisiaridis and Olivier Markowitch

QualSec Research Group, Departement d'Informatique, Université Libre de Bruxelles

## ABSTRACT

*Feature extraction and feature selection are the first tasks in pre-processing of input logs in order to detect cyber security threats and attacks while utilizing machine learning. When it comes to the analysis of heterogeneous data derived from different sources, these tasks are found to be time-consuming and difficult to be managed efficiently. In this paper, we present an approach for handling feature extraction and feature selection for security analytics of heterogeneous data derived from different network sensors. The approach is implemented in Apache Spark, using its python API, named pyspark.*

## 1. INTRODUCTION

Today, a perimeter-only security model in communication system is insufficient. With the Bring Your Own Device (BYOD) and IoT, data now move beyond the perimeter. For example, threats to the intellectual property and generally to sensitive data of an organization, are related either to insider attacks, outsider targeted attacks, combined forms of internal and external attacks or attacks, performed over a long period. Adversaries can be either criminal organizations, care- less employees, compromised employees, leaving employees or state-sponsored cyber espionage [6].

The augmentation of cyber security attacks during the last years emerges the need for automated traffic log analysis over a long period of time at every level of the enterprise or organizations information system. Unstructured, semi- structured or structured data in time-series with respect to security-related events from users, services and the underlying network infrastructure usually present a high level of large dimensionality and non-stationarity.

There is a plethora of examples in the literature as well as in open-source or commercial threat detection tools where machine learning algorithms are used to correlate events and to apply predictive analytics in the cyber security landscape.

Incident correlation refers to the process of comparing different events, often from multiple data sources in order to identify patterns and relationships enabling identification of events belonging to one attack or, indicative of broader malicious activity. It allows us to better understand the nature of an event, to reduce the workload needed to handle incidents, to automate the classification and forwarding of incidents only relevant to a particular consistency and to allow analysis to identify and reduce potential false positives.

Predictive Analytics, using pattern analysis, deals with the prediction of future events based on previously observed historical data, by applying methods such as Machine Learning [9]. For example, a supervised learning method can build a predictive model from training data to make predictions about new observations as it is presented in [7].

                                                                    

We need to build autonomous systems that could act in response to an attack in an early stage. Intelligent machines could implement algorithms designed to identify patterns and behaviors related to cyber threats in real time and provide an instantaneous response with respect to their reliability, privacy, trust and overall security policy framework.

By utilizing Artificial Intelligence (AI) techniques leveraged by machine learning and data mining methods, a learning engine enables the consumption of seemingly unrelated disparate datasets, to discover correlated patterns that result in consistent outcomes with respect to the access behavior of users, network devices and applications involved in risky abnormal actions, and thus reducing the amount of security noise and false positives. Machine learning algorithms can be used to examine, for example, statistical features or domain and IP reputation as it proposed in [1] and [12].

Along with history- and user-related data, network log data are exploited to identify abnormal behavior concerning targeted attacks against the underlying network infrastructure as well as attack forms such as man-in-the-middle and DDoS attacks [3] [9].

Data acquisition and data mining methods, with respect to different types of attacks such as targeted and indiscriminate attacks, provide a perspective of the threat landscape. It is crucial to extract and select the right data for our analysis, among the plethora of information produced daily by the information system of a company, enterprise or an organization [8]. Enhanced log data are then analyzed for new attack patterns and the outcome, e.g. in the form of behavioral risk scores and historical baseline profiles of normal behavior is forwarded to update the learning engine. Any unusual or suspected behavior can then be identified as an anomaly or an outlier in real or near real-time [11].

We propose an approach to automate the tasks of feature extraction and feature selection using machine learning methods, as the first stages of a modular approach for the detection and/or prediction of cybersecurity attacks. For the needs of our experiments we employed the Spark framework and more specifically its python API, *pyspark*.

Section 2 explains the difficulties of these tasks especially while working with heterogeneous data taken from different sources and of different formats. In section 3 we explain the difference between two common approaches in feature extraction by utilizing machine learning techniques. Section 4 deals with the task of extracting data from logs of increased data complexity. In section 5 we propose methods for the task of feature selection, while our conclusions are presented in section 6.

## 2. EXTRACTING AND SELECTING FEATURES FROM HETEROGENEOUS DATA

In our experiments, we examine the case where we have logs of records derived as the result of an integration of logs produced by different network tools and sensors (heterogeneous data from different resources). Each one of them monitors and records a view of the system in the form of records of different attributes and/or of different structure, implying thus an increased level of interoperability problems in a multi-level, multi-dimensional feature space; each network monitoring tool produces its own schema of attributes.

In such cases, it is typical that the number of attributes is not constant across the records, while the number of complex attributes varies as well [8]. On the other hand, there are attributes, e.g., dates, expressed in several formats, or other attributes referred to the same piece of information by using slightly different attribute names. Most of them are categorical, in a string format while the inner datatype varies from nested dictionaries, linked lists or arrays of further complex structure; each one of them may present its own multi-level structure which increases the level of complexity. In such cases, a clear strategy has to be followed for feature extraction. Therefore, we have to deal with flattening 1 and interoperability solving processes (Figure 1).

In our experiments, we used as input data an integrated log of recorded events produced by a number of different network tools, applied on a telco network. For the pre-processing analysis stage was used a single server of 2xCPUs, 8cores/CPU, 64GB RAM, running an *Apache Hadoop* installation v2.7 with Apache Spark v2.1.0 as a Standalone Cluster Mode, which can be regarded as a low-cost configuration for handling data exploration when dealing with huge amount of inputs; raw data volumes, for batch analysis, were approximately 16TBytes.

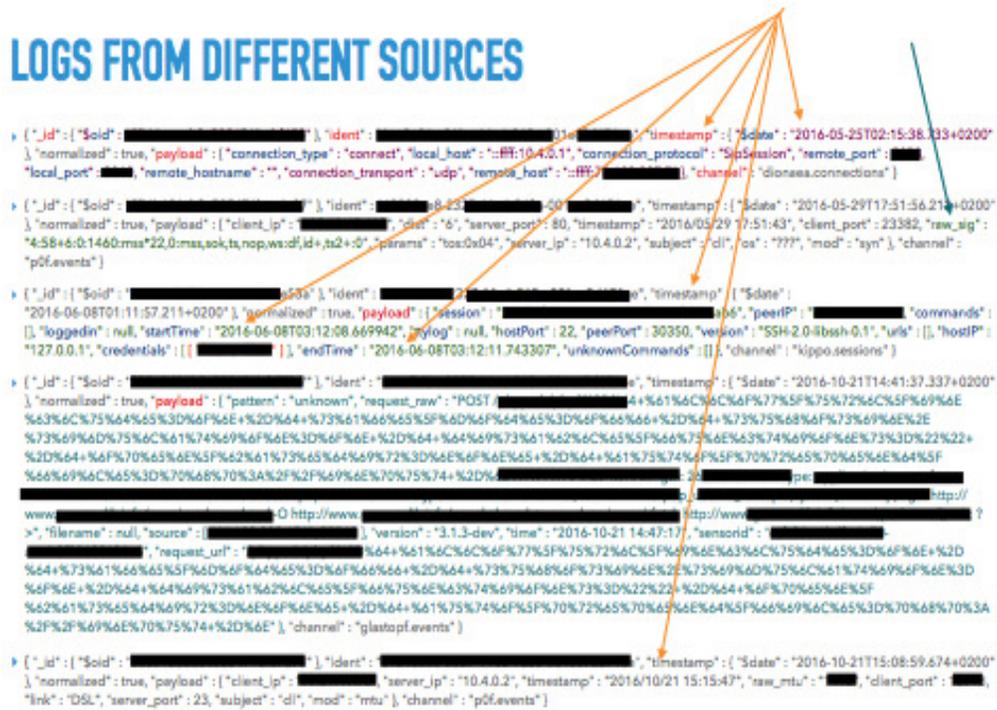

Figure 1.  Logs from different input sources

## 3. GLOBAL FLATTENING VS. LOCAL FLATTENING

The first question to be answered is related to the ability to define an optimal way to handle such complex inputs. Potential solutions may include:

- use the full number of dimensions (i.e. all the available features in each record), defined as global flattening
- decomposing initial logs into distinct baseline structures derived by each sensor/tool, defined as local flattening

### 3.1. LOOKING FOR THREATS AND ATTACKS IN A KILL CHAIN

In order to answer to these questions, we should also take into account the rationale behind the next steps. While working with the analysis of heterogeneous data taken from different sources, pre-process procedures, such as feature extraction, feature selection and feature transformation, need to be carefully designed in order not to miss any security-related significant events. These tasks are usually time-consuming producing thus significant delays to the overall time of the data analysis.

That is our main motivation in this work: to reduce the time needed for feature extraction in data exploration analysis by automating the process. In order to achieve it, we utilize the data model abstractions and we keep to a minimum any access to the actual data. The key characteristics of data inputs follow:

- logs derived from different sources
- heterogeneous data
- high-level of complexity
- information is usually hidden in multi-level complex structures

In the next stage, features will be transformed, indexed and scaled to overcome skewness, by following usually a normal distribution under a common metric space, in the form of vectors. As in our experiments we processed mainly un-labelled data (i.e. lack of any labels or any indication of a suspicious threat/attack), clustering techniques will be used to define baseline behavioural profiles and to detect outliers [1]. The latter may correspond to rare, sparse anomalies, that can be found by either first-class detection of novelties,n-gram analysis of nested attributes and pattern analysis using Indicators of Compromise (IoCs) [8] [6]. A survey on unsupervised learning outlier detection algorithms is presented in [14]. Finally, semi-supervised or/and supervised analysis can be further employed by using cluster labels, anomalous clusters, or experts feedback (using active learning methods), in order to detect and/or predict threats and attacks in near- and real-time analysis [3].

Outliers in time-series are expected to be found for a:

- single network sensor or pen-tester
- a subset of those, or
- by taking into account the complete available set of sensors and network monitoring tools

These time-series are defined in terms of either:

- time spaces as the contextual attributes
  - date attributes will be decomposed to time windows such as year, month, day of a week, hour and minute, following the approach proposed in [13] □
  - statistics will be calculated either for batch or online mode and then will be stored in HIVE tables, or in temporary views for ad-hoc temporal real-time analysis.
- a single time space (e.g. a specific day)
- a stable window time space (e.g. all days for a specific month)
- a user-defined variable window time space □

Our approach serves as an adaptation of the kill-chain model. The kill chain model [2] is an intelligence-driven, threat-focused approach to study intrusions from the adversaries' perspective. The fundamental element is the indicator which corresponds to any piece of information that can describe a threat or an attack. Indicators can be either atomic such as IP or email addresses, computed such as hash values or regular expressions, or behavioral which are collections of computed and atomic indicators such as statements.

Thus, in our proposal, contextual attributes represent either time-spaces in time-series, as the first level of interest, single attributes (e.g. a specific network protocol, or a user or any other atomic indicator), computed attributes (e.g. hash values or regular expressions), or even behavioral attributes of inner structure (e.g. collections of single and computed attributes in the form of statements or nested records). Then, outliers can be defined for multiple levels of interest for the remain behavioral attributes, by looking into single vector values [4], or by looking for the covariance and pairwise correlation (e.g. Pearson correlation) in a subset of the selected features or the complete set of features [5].

Experiments with data received from different network monitoring tools regarding the system of a telco enterprise, at the exploratory data stage revealed that the number of single feature attributes in this log were between a range of 7 (the smallest number of attributes of a distinct feature space) up to 99 attributes (corresponding to the total number of the overall available feature space). A fact, that led us to carry on with feature extraction by focusing on flattening multi-nested records separately for each different structure (under a number of 13 different baseline structures).

Thus, the main keys in the proposed approach for feature extraction are:

- extract the right data
  - correlation of the 'right data' can reveal long-term APTs □
  - re-usable patterns and trend lines as probabilities are indications of zero-day attacks
  - trend lines may also be used to detect DDoS attacks □
- handle interoperability issues □
- handle time inconsistencies, date formats, different names for the same piece □of information, by extending the NLP python library [2]

## 3.2. GLOBAL FLATTENING OF INPUT DATA

By following this approach, we achieve a full-view of entities behavior as each row is represented by the full set of dimensions. On the other hand, the majority of the columns do not have a value or it is set to Null. A candidate solution would be to use sparse vectors in the next stage of feature transformation, which in turn demands special care for NaN and Null values (for example, replace them either with the mean, the median, or with a special value). Most of the data in this stage are categorical. We need to convert them into numerical in the next stages, as in Spark, statistical analytics are available only for data in the form of a Vector or of the DoubleType.

This solution performs efficiently for a rather small number of dimensions while it suffers from the well-known phenomenon of the curse of dimensionality for a high number of dimensions, where data appear to be sparse and dissimilar in several ways, which prevents common data modelling strategies from being efficient.

## 3.3. LOCAL FLATTENING OF INPUT DATA

By following this approach, we identify all the different schemas in input data. First, it is a bottom-up analysis by re-sythesing results to answer to either simple of complex questions. In the same time, we can define hypotheses to the full set of our input data (i.e. top-down analysis) thus, it is a complete approach in data analytics, by allowing data to tell their story, in a concrete way, following a minimum number of steps. In this way, we are able to:

- keep the number of assumptions to a minimum
- look for misconfigurations and data correlations into the abstract dataframes definitions
- keep access to the actual data to a minimum
- provide solutions in interoperability problems, such as:
  - different representations of date attributes
  - namespace inconsistencies (e.g. attributes with names such as prot, protocol, connectionProtocol)
- cope with complex structures of different number of inner levels
- deal with event ordering and time-inconsistencies (as it is described in [10])

## 4. FEATURE EXTRACTION IN APACHE SPARK

In Apache Spark, data are organized in the form of dataframes, which resemble the well-known relational tables: there are columns (aka attributes or features or dimensions) and rows (i.e. events recorded, for example, by a network sensor, or a specific device). The list of columns and their corresponded datatypes define the schema of a dataframe. In each dataframe, its columns and rows i.e. its schema is unchangeable. Thus, an example of a schema could be the following:

DataFrame[id: string, @timestamp: string, honeypot: string, payloadCommand: string]

A sample of recorded events of this dataframe schema is shown in Figure 2:

Figure 2. A sample of recorded events, having 4 columns/dimensions

The following steps refer to the case in which logs/datasets are ingested in json format. Our approach examines the data structures on their top-level, focusing on abstract schemas and re-synthesis of previous and new dataframes, in an automatic way. Access to the actual data is only taken place when there is a need to find schemas in dictionaries and only by retrieving just one of the records (thus, even if we have a dataframe of million/billions of events, we only examine the schema of the first record/event). The words *field*, *attribute* or *column* refer to the same piece of data: a data frame column.

1) load the log file in a spark data frame, in json format
2) find and remove all single-valued attributes (this step applies also to the *feature selection* section)
3) flatten complex structures
   a) find and flatten all columns of complex structure (the steps are run recursively, down to the lowest complex attribute of the hierarchy of complex attributes)
      i) e.g. struct, nested dictionaries, linked lists, arrays, etc. (i.e. currently those which their value is of Row Type)
   b) remove all the original columns of complex structure
4) convert all time-columns into timestamps, using distinct time features in the data frames
5) integrate *similar* features in the list of data frames

Each attribute of a complex structure, such as of a *struct*, *nested dictionary*, *linked list* or an *array*, is handled in such way, which ensures that all single attributes of the lowest data level (i.e. the elements of an array, a dictionary, a list, etc.) will be flattened and expressed in 2-D.In this way, we manage to transform the schema of the original dataframe to a number of dataframes, each one corresponding to a schema that refers to a single network sensor or other input data source, as it is illustrated in the following figures (Figures 3, 4, 5 and 6). The steps, for these different cases follow:

- *struct* – RowType
  - use the leaf column at the last level of this struct-column to add it as a new column
- *list*: add list elements as new columns

- *array*: split array's elements and add the relevant new columns
- *dictionary* - steps:
  - find all inner schemas for attributes of type Dict, as a list
  - add the schemaType as an index to the original dataframe
  - create a list of dataframes, where each one has its own distinct schema
  - flatten all dictionary attributes, according to their schemas, in each dataframe of the list of dataframes by adding them as new columns

In Figure 3, in the left-hand schema*Schema#1*, attributes *_id* is of the datatype *struct*. The actual value is given by the inner-level attribute*, $oid*. The same stands for the outer attribute *timestamp*: the actual date value can be searched in the inner-level attribute *$date*. In both cases, attributes *$oid* and *$date* are extracted in the form of two new columns, named *_id_* and *dateOut*; the original attributes _id and timestamp are then deleted, having thus a new schema on the right-side, *Schema#2*. In this way, we achieved to reduce the complexity of the original input schema to a new one of lower complexity.

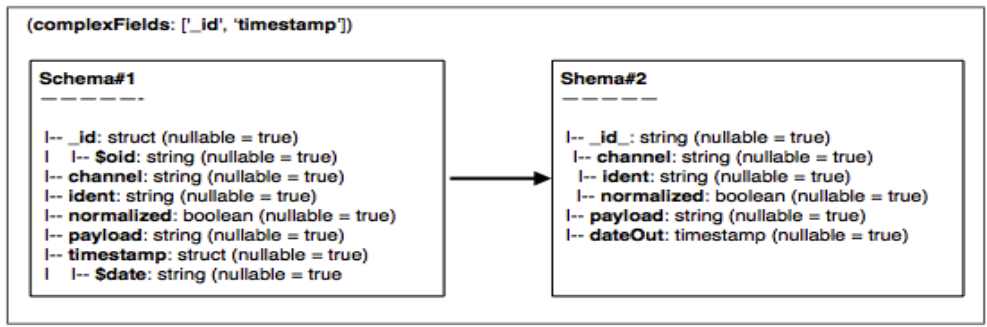

Figure 3. Transforming complex fields (i) – attributes **_id** and **timestamp** are of the datatype *struct*

In Figure 3, the exploratory analysis has revealed that the *payload* attribute represents actually a dictionary as a list of multi-nested dictionaries; each one of the latter present a complex structure with further levels. These different schemas found in payload are presented in Figure 4.

the 'Payload' column

['id', 'schemaType', 'remote_host', 'connection_protocol', 'local_port', 'connection_type', 'remote_hostname', 'remote_port', 'local_host', 'connection_transport', 'timestampDate', 'ident', 'channel']

['id', 'schemaType', 'client_ip', 'app', 'timestamp', 'server_ip', 'params', 'raw_sig', 'dist', 'client_port', 'mod', 'server_port', 'subject', 'timestampDate', 'ident', 'channel']

['id', 'schemaType', 'client_ip', 'server_ip', 'timestamp', 'uptime', 'subject', 'client_port', 'raw_freq', 'server_port', 'mod', 'timestampDate', 'ident', 'channel']

['id', 'schemaType', 'client_ip', 'server_ip', 'timestamp', 'reason', 'raw_hits', 'subject', 'client_port', 'mod', 'server_port', 'timestampDate', 'ident', 'channel']

['id', 'schemaType', 'client_ip', 'server_ip', 'timestamp', 'os', 'params', 'raw_sig', 'dist', 'client_port', 'mod', 'server_port', 'subject', 'timestampDate', 'ident', 'channel']

['id', 'schemaType', 'client_ip', 'server_ip', 'timestamp', 'link', 'subject', 'client_port', 'mod', 'server_port', 'raw_mtu', 'timestampDate', 'ident', 'channel']

['id', 'schemaType', 'hostIP', 'loggedin', 'commands', 'unknownCommands', 'startTime', 'peerPort', 'version', 'urls', 'session', 'ttylog', 'credentials', 'endTime', 'peerIP', 'hostPort', 'timestampDate', 'ident', 'channel']

['id', 'schemaType', 'sensorid', 'request_raw', 'request_url', 'filename', 'source', 'pattern', 'version', 'time', 'timestampDate', 'ident', 'channel']

['id', 'schemaType', 'tos', 'ttl', 'ethdst', 'ethtype', 'udplength', 'sensor', 'priority', 'destination_ip', 'timestamp', 'signature', 'classification', 'ethlen', 'dgmlen', 'destination_port', 'header', 'source_port', 'proto', 'source_ip', 'iplen', 'ethsrc', 'timestampDate', 'ident', 'channel']

['id', 'schemaType', 'destination_port', 'timestamp', 'tcpflags', 'tcpwin', 'dgmlen', 'tcpack', 'classification', 'sensor', 'proto', 'tcpseq', 'header', 'source_ip', 'iplen', 'tos', 'ttl', 'ethtype', 'priority', 'destination_ip', 'tcplen', 'ethlen', 'ethdst', 'source_port', 'signature', 'ethsrc', 'timestampDate', 'ident', 'channel']

['id', 'schemaType', 'timestamp', 'destination_ip', 'dgmlen', 'classification', 'sensor', 'proto', 'header', 'source_ip', 'iplen', 'tos', 'ttl', 'ethtype', 'priority', 'icmpcode', 'icmpseq', 'ethlen', 'ethsrc', 'ethdst', 'icmpid', 'signature', 'icmptype', 'timestampDate', 'ident', 'channel']

['id', 'schemaType', 'daddr', 'md5', 'url', 'dport', 'sport', 'sha512', 'saddr', 'timestampDate', 'ident', 'channel']

['id', 'schemaType', 'url', '@timestamp', 'honeypot', 'payloadCommand', 'headers', 'method', 'payloadMd5', 'form', 'payloadBinary', 'payloadResource', 'type', 'source', 'timestampDate', 'ident', 'channel']

Figure 4. Different schemas in the **payload** attribute

In Figure 5, we illustrate the new-created dataframes schemas which correspond to the different schemas of the payload attribute in Figure 4. By following this approach, data are easier to be handled: in the next stages, they will be cleaned, transformed from categorical to numerical and then they will be further analyzed in order to detect anomalies in entities behavior.

The dataframe schema in Figure 6 is the second of the dataframes derived by flattening the *payload* attribute (Figure 5) into its inner-level attributes. Here, feature *raw_sig* is in the form of an array. By applying consecutive transformations automatically, we manage to extract all inner attributes, which simplifies the process of correlating data in the next stage. Thus, by looking into the *raw_sig* column, we identify inner values separated by '**:**', which further are decomposed into new features derived by the inner levels, as it is depicted e.g. for column *attsCol5*; the latter could be further split by leading to two new columns (e.g. with values *1024* and *0*, respectively), as this process is recursive and automated; special care is given how we name the new columns, in order to follow the different paths of attributes decomposition.

```
'0' : DataFrame[channel: string, ident: string, normalized: boolean, _id_: string, schemaType: string,
remote_host: string, connection_protocol: string, local_port: string, connection_type: string,
remote_hostname: string, remote_port: string, local_host: string, connection_transport: string, dateOutIn:
timestamp]

'1' : DataFrame[channel: string, ident: string, normalized: boolean, _id_: string, schemaType: string,
client_ip: string, app: string, server_ip: string, params: string, raw_sig: string, dist: string, client_port: string,
mod: string, server_port: string, subject: string, dateOutIn: timestamp, timestampIn: timestamp]

'2' '' : DataFrame[channel: string, ident: string, normalized: boolean, _id_: string, schemaType: string,
client_ip: string, server_ip: string, uptime: string, subject: string, client_port: string, raw_freq: string,
server_port: string, mod: string, dateOutIn: timestamp, timestampIn: timestamp]

'3' : DataFrame[channel: string, ident: string, normalized: boolean, _id_: string, schemaType: string,
client_ip: string, server_ip: string, reason: string, raw_hits: string, subject: string, client_port: string, mod:
string, server_port: string, dateOutIn: timestamp, timestampIn: timestamp]

'4' : DataFrame[channel: string, ident: string, normalized: boolean, _id_: string, schemaType: string,
client_ip: string, server_ip: string, os: string, params: string, raw_sig: string, dist: string, client_port: string,
mod: string, server_port: string, subject: string, dateOutIn: timestamp, timestampIn: timestamp]

'5' : DataFrame[channel: string, ident: string, normalized: boolean, _id_: string, schemaType: string,
client_ip: string, server_ip: string, link: string, subject: string, client_port: string, mod: string, server_port:
string, raw_mtu: string, dateOutIn: timestamp, timestampIn: timestamp]

'6' : DataFrame[channel: string, ident: string, normalized: boolean, _id_: string, schemaType: string, hostIP:
string, loggedin: string, commands: string, unknownCommands: string, peerPort: string, version: string, urls:
string, session: string, ttylog: string, credentials: string, peerIP: string, hostPort: string, dateOutIn: timestamp,
startTimeIn: timestamp, endTimeIn: timestamp]

'7' : DataFrame[channel: string, ident: string, normalized: boolean, _id_: string, schemaType: string,
sensorid: string, request_raw: string, request_url: string, filename: string, source: string, pattern: string,
version: string, dateOutIn: timestamp, timeIn: timestamp]

'8' : DataFrame[channel: string, ident: string, normalized: boolean, _id_: string, schemaType: string, tos:
string, ttl: string, ethdst: string, ethtype: string, udplength: string, sensor: string, priority: string,
destination_ip: string, signature: string, classification: string, id: string, ethlen: string, dgmlen: string,
destination_port: string, header: string, source_port: string, proto: string, source_ip: string, iplen: string,
ethsrc: string, dateOutIn: timestamp, timestampIn: timestamp]

'9' : DataFrame[channel: string, ident: string, normalized: boolean, _id_: string, schemaType: string,
destination_port: string, tcpflags: string, tcpwin: string, dgmlen: string, tcpack: string, classification: string,
sensor: string, proto: string, tcpseq: string, header: string, priority: string, iplen: string, tos: string, ttl:
string, ethtype: string, priority: string, destination_ip: string, id: string, tcplen: string, ethlen: string, ethdst:
string, source_port: string, signature: string, ethsrc: string, dateOutIn: timestamp, timestampIn: timestamp]

'10 : DataFrame[channel: string, ident: string, normalized: boolean, _id_: string, schemaType: string,
destination_ip: string, dgmlen: string, classification: string, sensor: string, proto: string, header: string,
source_ip: string, iplen: string, tos: string, ttl: string, ethtype: string, priority: string, icmpcode: string, id:
string, icmpseq: string, ethlen: string, ethsrc: string, ethdst: string, icmpid: string, signature: string, icmptype:
string, dateOutIn: timestamp, timestampIn: timestamp]

'11' : DataFrame[channel: string, ident: string, normalized: boolean, _id_: string, schemaType: string, daddr:
string, md5: string, url: string, dport: string, sport: string, sha512: string, saddr: string, dateOutIn: timestamp]

'12' : DataFrame[channel: string, ident: string, normalized: boolean, _id_: string, schemaType: string, url:
string, honeypot: string, payloadCommand: string, headers: string, method: string, payloadMd5: string, form:
string, payloadBinary: string, payloadResource: string, type: string, source: string, dateOutIn: timestamp,
@timestampIn: timestamp]
```

Figure 5. The new-created dataframes which correspond to the different schemas in **payload**

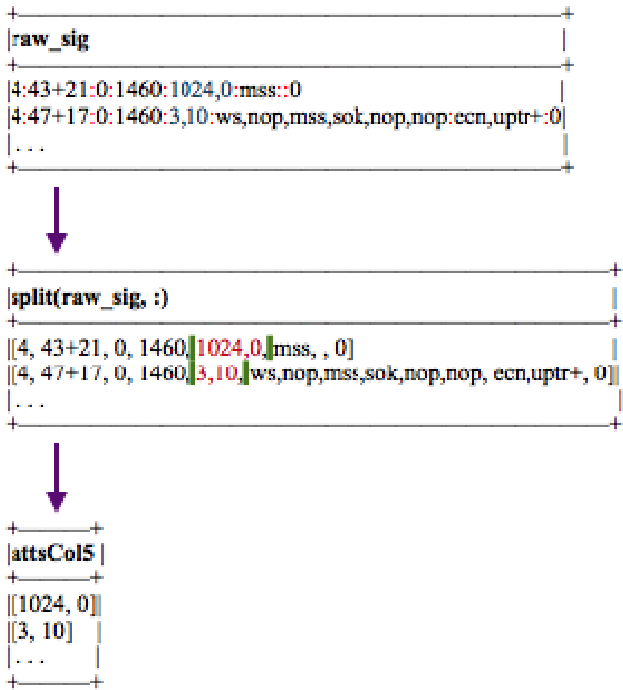

Figure 6. Transforming complex fields (iii) – attributes ***raw_sig*** is of the datatype *array*

## 5. FEATURE SELECTION

The process of feature selection (FS) is crucial for the next analysis steps. As was explained in 3.1, our motivation in our approach is to reduce data complexity in parallel with a significant reduction of the time needed for applying security analytics in un-labelled data. As we are aiming ultimately to detect anomalies as strong form of outliers in order to improve quantitative metrics such as, to increase accuracy and detection rates or to decrease security noise to a minimum, we need to select the data that are more related to our questions. Dimensionality reduction can play a significant role in complex event processing, especially when data are coming from different sources and different forms.

We present four methods to achieve this goal:

1. Leave-out single-value attributes
2. Namespace correlation
3. Data correlation using the actual values
4. FS in case of having a relative small number of categories

### 5.1. LEAVE-OUT SINGLE-VALUE ATTRIBUTES

The first method is quite simple: all single-valued attributes are removed from the original dataframe. For example, consider the dataframe schema in Figure 7. Attribute *normalized* of datatype *Boolean* takes the value *True* for all the events in our integrated log and therefore we drop the relevant column, which leads to a new dataframe schema.

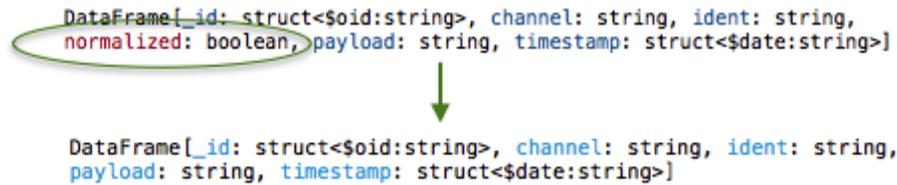

Figure 7. Attribute **normalized** is left out as it presents a constant value of *True* in all records

## 5.2. NAMESPACE CORRELATION

It is quite common when data inputs are coming from different sources to deal with entity attributes which refer to the same piece of information although their names are slightly different. For example, attributes *proto* and *connection protocol* refer to the actual protocol used in a communication channel. Different tools used by experts to monitor network traffic do not follow a unified namespace scheme. This fact, could lead to misinterpretations, information redundancy and misconfigurations in data modelling, among other obstacles in data exploration stage; all these refer mainly to problems in interoperability, as can be seen in Figure 8. By solving such inconsistencies, we achieve to further reduce data complexity as well as to reduce the overall time for data analysis. In [10] we have presented an approach to handle such interoperability issues by utilizing means derived by the theory of categories.

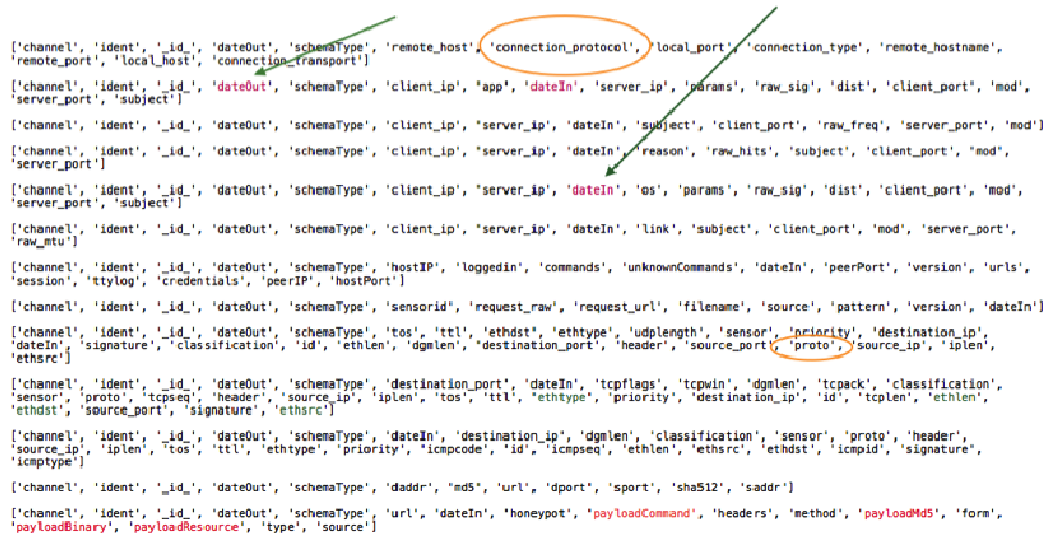

Figure 8. Attributes **proto** and **connection_protocol** refer to the same piece of data

## 5.3. USING PEARSON CORRELATION TO REDUCE THE NUMBER OF DIMENSIONS

As long as data inputs, in the form of dataframes, are cleaned, transformed, indexed and scaled into their corresponding numerical values, and before the process of forming the actual feature vectors that will be used in clustering, by using data correlation, we are able to achieve a further reduction of the dimensions that will be used for the actual security analytics.

The outcome of applying this technique, using *Pearson correlation*, is presented in Figure 9. Attributes highly correlated may be omitted while defining the relevant clusters; the choice of the particular attribute to be left out is strongly related to the actual research of interest. For example, we are interested to monitor the behavior in local hosts and to detect any anomalies deviate by patterns of normal behavior.

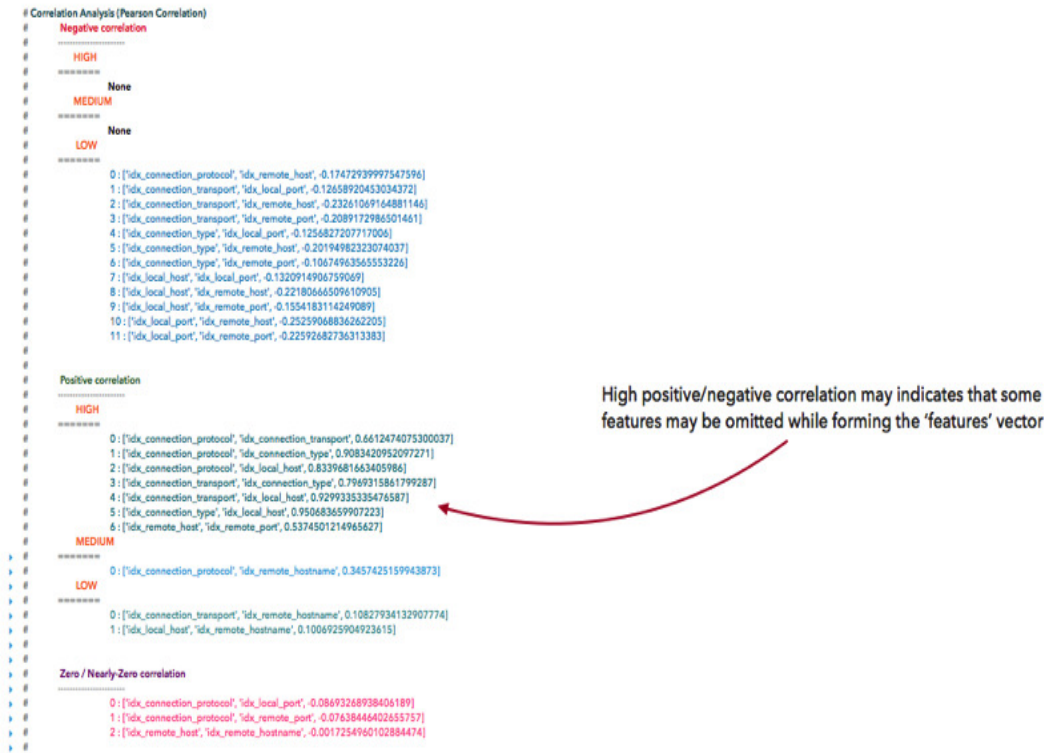

Figure 9. Applying Pearson correlation to indexed and scaled data for feature selection

## 5.4. Feature Selection In Case Of Having A Relative Small Number Of Categories

In case where we deal with categorical attributes presenting a relative small number of categories, i.e. *numberOfCategories<= 4*, we propose the following steps in order to achieve a further feature reduction. We distinguish the cases where data are un-labelled (lack of any indication for a security-related event) and the case where some or all the labels are available. We need to mention that in real scenarios, usually we need to cope with either fully un-labelled data or highly-unbalanced data (i.e. where only few instances of the rare/anomalous class are available). Working with un-labelled data:

- For the set of these features, select each one of them as the feature-label attribute and then either:
- Use a decision tree with a *multi-class classification evaluator* to further reduce the number of the dimensions (by following one or more of the aforementioned techniques)
- Create $2^n$ sub-dataframes with respect to the number of categories
- Calculate features importance using a *Random Forest classifier*
- Use an ensemble technique in the form of a *Combiner* e.g. a *neural network* or a *Bayes classifier*, running a combination of the above techniques to optimize results in the next levels of the analysis (e.g. to further optimize detection rates)

Working with labelled data:

- Select features using the *Chi-Square test* of independencies. In our experiments, with respect to the input data, we have used four different statistical methods, available in *Spark ML library*:
  - The number of top features
  - A fraction of the top features
  - p-values below a threshold to control the false positive rate
  - p-values with false discovery rate below a threshold

## 6. CONCLUSIONS

We have presented an approach to handle efficiently the tasks of feature extraction and feature selection while working with security analytics by utilizing machine learning techniques. It is an automated solution to handle interoperability problems. It is based on a continuous transformation of the abstract definitions of the data inputs.

In the case of feature extraction, access to the actual data is limited to a minimum read actions of the first record of a dataframe and only when it is needed to extract the inner schema of a dictionary-based attribute. In the case of feature selection, the actual data are accessed only to find correlations between them, before we apply clustering or any other method for threat detection.

By following the proposed approach, we managed to achieve our primary objectives: reduce computational time, reduce data complexity and provide solutions to interoperability issues, while analyzing vast amount of heterogeneous data from different sources.
The approach can be formalized in the next steps by utilizing novel structures derived from the theory of categories as it has been presented in [10] towards an overall optimization, in terms of quantitative and qualitative metrics.

## AUTHORS

Dimitrios Sisiaridis received his BSc in Information Engineering by ATEI of Thessaloniki. He received his MSc in Computing (advanced databases) and PhD in applied maths and security by Northumbria University in Newcastle. He is a member of the Qualsec Research Group, in Université Libre de Bruxelles, working as a researcher in projects related to big data security analytics

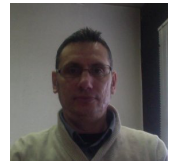

Prof. Olivier Markowitch is associated Professor of the Departement d' Informatique, in Université Libre de Bruxelles. He is a member of the QualSec Research Group, working on the design and analysis of security protocols and digital signature schemes

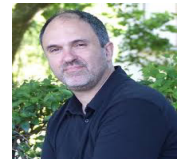